\documentclass[twocolumn,prd,nofootinbib,showpacs]{revtex4}
\usepackage{graphicx}

\newcommand{\be}{\begin{equation}}
\newcommand{\ee}{\end{equation}}
\newcommand{\bea}{\begin{eqnarray}}
\newcommand{\eea}{\end{eqnarray}}

\begin{document}
\title{``Discrepant hardenings'' in cosmic ray spectra: a first estimate of the effects on secondary antiproton and diffuse gamma-ray yields.}
\author{Fiorenza Donato}\email{donato@to.infn.it}
\affiliation{Dipartimento di Fisica Teorica, Universit\`a di Torino
and INFN--Sezione di Torino, Via P. Giuria 1, 10122 Torino, Italy}
\author{Pasquale D.~Serpico}\email{serpico@lapp.in2p3.fr}
\affiliation{LAPTh, UMR 5108, 9 chemin de Bellevue - BP 110, 74941 Annecy-Le-Vieux, France}

\begin{abstract}
Recent data from CREAM seem to confirm early suggestions that primary  cosmic ray (CR) spectra at few TeV/nucleon are harder than in the 
10-100 GeV range. Also, helium and heavier nuclei spectra appear systematically harder than the proton fluxes at corresponding energies.
We note here that if the measurements reflect intrinsic features in the interstellar fluxes (as opposed to local effects) appreciable modifications 
are expected in the sub-TeV range for the secondary yields, such as antiprotons and diffuse gamma-rays. Presently, the ignorance on
the origin of the features represents a systematic error in the extraction of astrophysical parameters as well as for background estimates for
indirect dark matter searches. We find that the spectral modifications are appreciable above 100 GeV, and can be responsible for  $\sim$30\% effects for antiprotons at 
energies close to 1 TeV or for gamma's at energies close to 300 GeV, compared to currently considered predictions based on simple extrapolation of input 
fluxes from low energy data.  
Alternatively, if the feature originates from local sources, uncorrelated spectral changes might show up in antiproton and high-energy gamma-rays, with the latter ones likely dependent from the line-of-sight.
\end{abstract}
\pacs{98.70.Sa\hfill LAPTH-043/10}
\maketitle

\section{Introduction}
A more accurate determination of primary cosmic ray spectra at the top of the atmosphere has obvious  implications for
the understanding of the acceleration and propagation of Galactic cosmic rays.  It is also crucial for other fields of
investigations in astroparticle physics, two notable examples being atmospheric neutrino studies 
(e.g.~\cite{Barr:2006it}) and the calculation
of the backgrounds for indirect dark matter (DM)
searches (see for example~\cite{Donato:2001ms}).

In the specific case of indirect DM searches, an important {\it implicit} assumption is that fluxes 
measured at the top of the atmosphere, at 
sufficiently high energies to avoid solar modulation effects, are representative of interstellar medium (ISM) spectra. Or, more correctly, one often assumes
universality for the injection term and the propagation properties, fitting the free parameters (like injection and diffusion power-law index) in such a way to reproduce the observed spectra.
It is those ``universal'' interstellar spectra which in turn enter as source term of secondary yields (like antiprotons or diffuse gamma rays) due to inelastic 
collisions in the ISM. This assumption is usually supported by the apparent featureless nature of the observed cosmic ray fluxes (suggesting, at least in average, some universal
mechanism for production and propagation) as well as by the check a posteriori that the diffuse gamma-ray radiation of hadronic origin has a spectrum consistent
with the hypothesis, within the errors.
However, especially at energies larger than the TeV scale,  inferring accurate spectra is challenging due the scarce statistics and experimental difficulties, making the above arguments at best based on shaky observational evidence.
Also, for assessing uncertainties in secondary yields, the usual practice is to fit primary data  to some power-law parameterization and extrapolate to 
high energies. While this is a reasonable prescription for most applications given the present level of understanding, these 
simplified approaches and assumptions might hide a systematic error when searching for signatures showing peculiar energy features. For example, for
antiprotons this is the case involving contributions from DM annihilation~\cite{Donato:2008jk} or production at the
sources~\cite{Blasi:2009bd}.

Obviously, the standard prescriptions do not usually account for the possibility that a {\it systematic} departure (rather than
statistical scattering) is present in  the {\it spectral shape} of  the fitting formula, which is mostly calibrated on low energy
data, neither of the possibility that observed spectra might not be fair representative of the interstellar ones.  Recent data from the 
CREAM balloon-borne experiment~\cite{Ahn:2010gv} seem to confirm earlier suggestions (see
e.g.~\cite{Panov:2006kf}) that cosmic ray spectra at few  TeV/nucleon are harder than in the 10-100 GeV range, and that helium (He) and
heavier nuclei fluxes are harder than the proton ($p$) flux at corresponding energies. Preliminary data from PAMELA also suggest a
hardening in $p$ and He spectra at a rigidity of about 250 GV, with a He spectrum having an index $\sim 0.1$ lower than the proton one over
all energies above a few GeV~\cite{adrianitalk}.  All this motivated us to have a more careful look at the errors potentially
committed when estimating secondary yields in the interstellar medium.  

In this article, we refrain from discussing possible astrophysical interpretations of the above mentioned features, although some have been proposed, see~\cite{Ahn:2010gv,Biermann:2010qn}. 
We note however that if the measurements reflect intrinsic properties of the interstellar spectra, 
appreciable modifications (i.e. above $\sim 10\%$) of very specific spectral shape are expected for the secondary yields  in the 0.1 to 1 TeV range,
which is directly accessible (with growing precision) to present and forthcoming experiments like PAMELA, FERMI, and AMS-02. On the other
hand, if the hardenings reflect {\it local} phenomena/sources, secondary yields which probe a large volume of the ISM, like antiprotons, might not show relevant departures from 
naive expectations, while the diffuse gamma-rays along different lines of sight might reveal different hardenings reflecting the primary spectra present in different
regions of the ISM. Clearly, this provides
an important test for theories about the origin of the breaks. 
To the best of our knowledge, present data in high
energy astrophysics are either unrelated to the hardenings discussed here or still  of too limited precision to provide a crucial test, but the situation
is likely to change in the near future.

This article is structured as follows: in Sec.~\ref{input} we discuss the input fluxes and parameterization used to provide a first estimate of the effect. In Sec.~\ref{results}
we present the results for antiprotons and $\gamma$-rays, finally in Sec.~\ref{disconcl} we discuss some implications of our findings, and conclude. 

\section{Input fluxes}~\label{input}
In the present exploratory study, we refrain from the ambitious goal of analyzing the whole body of cosmic ray flux data in the
$10-10^4\,$GeV/n range. Rather we limit ourselves to provide a first assessment of the systematic effect potentially introduced by deviations
from the power law behaviour at high energy, in general with different spectral indexes for different species. To this purpose, we explore the
effects of combining the fits of ``low-energy'' (namely in the range about 10-100 GeV/n) 
 proton ($i=1$) and helium ($i=2$) flux 
data, $\phi_i^L$,  taken from AMS-01~\cite{Aguilar:2002ad} 
(in turn, to large extent consistent with what reported by other experiments), 
  with the ``high-energy'' (above about 1 TeV/n) fluxes $\phi_i^H$ inferred by
CREAM~\cite{Ahn:2010gv}. We adopt broken power-laws to connect the two sets, 
using the following flux parameterizations (differential fluxes
with respect to kinetic energy per nucleon $T$):
\begin{eqnarray}
\phi_1(T)&=&\phi_{1}^L(T)\Theta(B_1-T)+\phi_{1}^{H}(T)\Theta(T-B_1)\,,\label{broknP}\\
\phi_2(T)&=&\phi_{2}^L(T)\Theta(B_2-T)+\phi_{2}^{H}(T)\Theta(T-B_2)\,.
\label{eq:fit}
\end{eqnarray}

The fluxes ``$L$'' are the best fit values taken from AMS-01~\cite{Aguilar:2002ad},  
rewritten in terms of kinetic energy $T$ per nucleon (in GeV/n) instead of 
rigidity and asymptotically decreasing as $\sim T^{-2.78}$ for $p$ and $T^{-2.74}$ for He. 
In units of $({\rm GeV/n\,m^2\,s\,sr})^{-1}$, they write
\begin{eqnarray}
\phi_{1}^{L}(T)&=&1.71\times 10^4\,\left(\sqrt{(T+m_p)^2-m_p^2}\right)^{-2.78}\,,\label{amsP}\\
\phi_{2}^{L}(T)&=&5.04\times 10^3\,\left(\frac{\sqrt{(4\,T+m_{\rm He})^2-m_{\rm He}^2}}{{2}}\right)^{-2.74}\label{amsHe}\,.
\end{eqnarray}
The high energy fluxes ``$H$'' are taken from CREAM with the following criteria: 
i) power-laws in $T$ are assumed, with the spectral indexes fixed to the best-fit values reported 
in~\cite{Ahn:2010gv}, i.e. 2.66 for $p$ and 2.58 for He; 
ii)  the proton spectrum normalization is taken from the first CREAM point in Fig. 3 of~\cite{Ahn:2010gv}; 
iii) the Helium spectrum normalization follows from imposing that at $T=9$ TeV/nucleon the proton to helium flux ratio 
is equal to 8.9~\cite{Ahn:2010gv}. 
As a result, in units of $({\rm GeV/n\,m^2\,s\,sr})^{-1}$,
 \begin{eqnarray}
\phi_{1}^{H}(T)&=&7.42\times 10^3\,T^{-2.66}\,,\\
\phi_{2}^{H}(T)&=&
4.03\times 10^{2}\,T^{-2.58}\,.
\end{eqnarray} 
The crossover energies $B_1,\,B_2$  for the broken power-laws are simply obtained by continuity, and are approximately $B_p=1000\,{\rm GeV}$, $B_{\rm He}=30\,{\rm GeV}$/n 
for the parameters above\footnote{Assuming a {\it relative} uncertainty in the flux normalization of the two experiments of $\alt 20\%$---certainly consistent with published values---would suffice to 
bring these crossover values in consistency with the rigidity $\sim 250$ GV hinted to by PAMELA, ref.~\cite{adrianitalk}}.
A comparison with the predictions following from the extrapolation of the AMS-01 fits (i.e. the $\phi_i^L$ of Eqs.(\ref{amsP},\ref{amsHe})) 
to arbitrarily high energy will be presented to provide an estimate of the impact of high-energy spectral uncertainty on the secondary yield flux.

\section{Results}\label{results}
Discrepant hardenings of primary cosmic ray fluxes, possibly of non-universal nature, would obviously affect all
the yields of $e^{+},\, \bar{p}$ and $\gamma$ secondaries produced by collisions in the interstellar medium (ISM). 
Here we do not discuss  charged leptons simply because the primary 
flux effects do not provide the major uncertainty in the flux shape (even fixing the average propagation parameters): 
very likely recent data~\cite{Adriani:2008zr,Abdo:2009zk,Ackermann:2010ij}  indicate that additional sources of ``primary'' 
positrons exist  for which the above mentioned effects are  expected to be sub-leading (see e.g.~\cite{Serpico:2008te}). 
Additionally, energy losses make the range shorter and the computation of the actual flux at the Earth non-trivial, 
so it would be more difficult to disentangle the effects due to the break in primary spectra from a complicated interplay of 
effects involving the discreteness of local sources, inhomogeneities in the radiation field, etc. as illustrated for instance 
in~\cite{Delahaye:2010ji}.
\begin{figure}[t]
\vspace{-30.0mm}
\includegraphics[angle=0,width=0.55\textwidth]{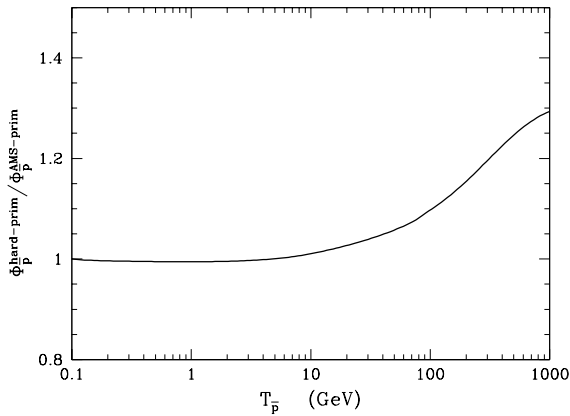} 
\vspace{-1cm}
\caption{Ratio of antiproton fluxes from hard sources (Eqs.(\ref{broknP}, \ref{eq:fit})) to 
the same flux obtained with p and He extrapolated from AMS data 
to all energies (see text for details).
~\label{fig:ap1}}
\end{figure} 
The effect of universal primary CR hardening should be appreciable in the predicted shape of the antiproton or diffuse gamma-ray signal. 
Here we report a
careful computation of the effect on the antiproton spectrum, where the impact is expected to be the largest in view of future
high-statistics results from AMS-02,  and an estimate of the effect on the hadronic gamma-ray diffuse background, of some interest for the
interpretation of FERMI data.

\subsection{Effects on secondary antiprotons.} 
The computation of the secondary $\bar{p}$ flux has been performed as described in 
Refs.~\cite{Donato:2008jk,Donato:2001ms}, to which we refer for all the details. The only 
component which we will modify in the present calculation is the input $p$ and He spectra. 
We briefly remind that secondary $\bar{p}$ are yielded by the spallation of cosmic ray 
proton and helium nuclei over the H and He nuclei in the ISM,  the contribution of heavier nuclei being 
negligible.
The framework used to calculate the antiproton flux is a two--zone
diffusion model with convection and reacceleration, as well as 
spallations on the ISM, electromagnetic energy losses and the so--called 
tertiary component, corresponding to non--annihilating inelastic scatterings
on the ISM. 
The relevant transport parameters 
are constrained from the boron-to-carbon (B/C) analysis \cite{usineI} and correspond to:
i) the half thickness of the diffusive halo
of the Galaxy $L$; ii) the normalization of the diffusion coefficient $K_0$ and
its slope $\delta$ ($K(E)=K_0 \beta R^\delta$);
iii) the velocity of the constant wind directed perpendicular to the galactic disk
$\vec{V_c} = \pm V_c \vec{e_z}$; and iv) the reacceleration intensity parameterized 
by the the Alfv\'enic speed $V_a$. 
The above parameters show significant degeneracies when confronted to B/C data
\cite{usineI}. Nevertheless, the impact on the secondary $\bar{p}$  flux 
is marginal \cite{Donato:2001ms}. 
The fluxes presented below have been obtained for the B/C  best fit propagation parameters,
i.e. $L = 4$~kpc, $K_0 = 0.0112$~kpc$^2$Myr$^{-1}$, $\delta=0.7$,
$V_c = 12.$~km~s$^{-1}$ and $V_a = 52.9$~km~s$^{-1}$ \cite{usineI}. 

We are interested in the effect of primary $p$ and He hardening at high energies 
on the $\bar{p}$ flux and therefore concentrate on the relative shape
effect through  antiproton flux ratios. Our results are reported in Fig.~\ref{fig:ap1}, 
where we plot the ratio of antiproton fluxes  obtained with two different primary spectra.
The flux at the numerator has been obtained with 
the spectra in Eqs.(\ref{broknP}, \ref{eq:fit}), while in the denominator we employ the fit 
to AMS data arbitrarily extrapolated to the highest energies. 
The modification of the antiproton flux clearly reflects in its shape.
The effect of the hardening of primary spectra at hundreds of GeV/n 
starts to be visible on the antiproton flux at around 100 GeV. It 
is near 15\% at 200 GeV and reaches 30\% at 1 TeV.  
Given the weak dependence of the secondary antiproton flux on the B/C selected 
transport parameters, our results can be considered nearly independent
of the propagation model.
If the hardening of primary nuclei will be confirmed at high energies, 
a spectral distortion of the secondary antiproton flux has to be expected.
This effect could be potentially observable by a future high 
precision space-based mission, such as AMS-02.

\subsection{Effects on hadronic diffuse gamma-rays.}\label{hadg}
\begin{figure}[t]
\begin{center}
\begin{tabular}{c}
\includegraphics[angle=0,width=0.5\textwidth]{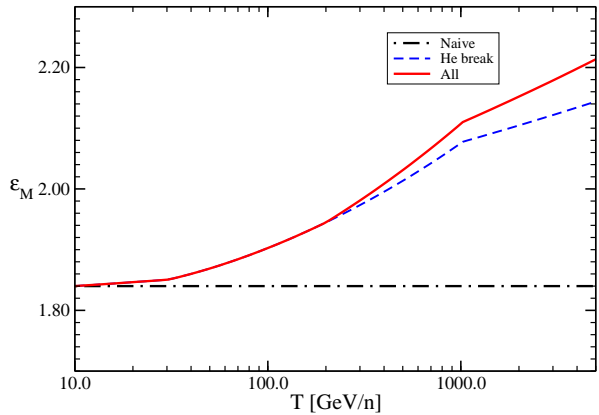} 
\end{tabular}
\end{center}
\caption{Enhancement factor, see Eqs.~(\ref{eps0},\ref{epsN}), for the three representative cases described in Sec.~\ref{hadg}.~\label{fig:1}}
\end{figure}

The cosmic gamma ray flux observed in our Galaxy is expected to be mainly due to the inelastic scattering of 
incoming CRs on the nuclei of the ISM. The involved hadronic reactions produce gamma rays mostly via $\pi^0$ decays. In addition to 
this hadronic component, other contributions are expected ---at different levels depending on the specific model--- to Inverse Compton 
and bremsstrahlung radiation.   
The basic models for the production of gamma rays from  $\pi^0$ decays, 
considered for example by the Fermi-LAT Collaboration~\cite{Abdo:2009ka}, do not introduce high-energy 
spectral breaks in the proton spectrum $\phi_1$, and account for  nuclear effects (both in CR spectra and in target composition) 
in the $\pi^0$ yield simply by rescaling the $pp$ production via a {\it constant} ``nuclear enhancement factor'', 
taken from the value at the reference energy $T_{*}\equiv$10 GeV/n reported in~\cite{Mori:2009te}. 
This enhancement encodes  the relative yield of gamma-rays from nucleus-$p$ and nucleus-Helium collisions compared 
with that from $p$-$p$ collisions via appropriate factors $m_{ip}\,,\,m_{i\alpha}$,  basically constant at $T>10\,$GeV/n 
(the effects discussed in this paper are only relevant at high energy, so it's enough to focus on quantities at $T>10\,$GeV/n).  
This enhancement is defined as
\begin{equation}
\epsilon_{\rm M}(T)=\sum_i m_{i1}{\phi_i(T) \over \phi_1(T)}
+\sum_i m_{i2}{\phi_i(T) \over \phi_1(T)}\label{eps0}
\times {r \over 1-r}\,,
\end{equation}
where the index $i$ runs over all CR species (including protons, $i=1$), $r\simeq 0.096$ is the He/H fraction in the ISM and $\phi_i$ being the CR spectrum of the species $i$. If all the nuclei have roughly identical $T$ dependence of their spectra, as suggested in~\cite{Ahn:2010gv}, one can write
\begin{equation}
\epsilon_{\rm M}(T)=1+{m_{12}\,r \over 1-r}+\left(m_{21}+{m_{22}\,r \over 1-r}\right){\phi_2 \over \phi_1}+
k_N\,{\phi_N\over \phi_1}\,,\label{epsN}
\end{equation}
 where $\phi_N(T)$ is any nuclear-like CR flux, and $k_N$ is a normalization factor. In Fig.~\ref{fig:1} we show $\epsilon_{\rm M}(T)$ for three
 cases: i) the constant value $\epsilon_{\rm M}=1.84$, adopted for example in~\cite{Abdo:2009ka} (long-dashed, black);
 ii) the fluxes of $p$ and He are set to the broken power-law functions described above, while the last term $k_N\,\phi_N/\phi_1$ is taken constant in
 energy and fixed so that  $\epsilon_{\rm M}(T_{*})=1.84$ (short-dashed, blue). iii) As in ii) for $p$ and He, but assuming for nuclei heavier than He
 a constant contribution to $\epsilon$ below $200\,$GeV/n (so that  $\epsilon_{\rm M}(T_{*})=1.84$),  then rising as $T^{0.1}$, as
 suggested by CREAM data (solid, red).

\begin{figure}[t]
\begin{center}
\begin{tabular}{c}
\includegraphics[angle=0,width=0.5\textwidth]{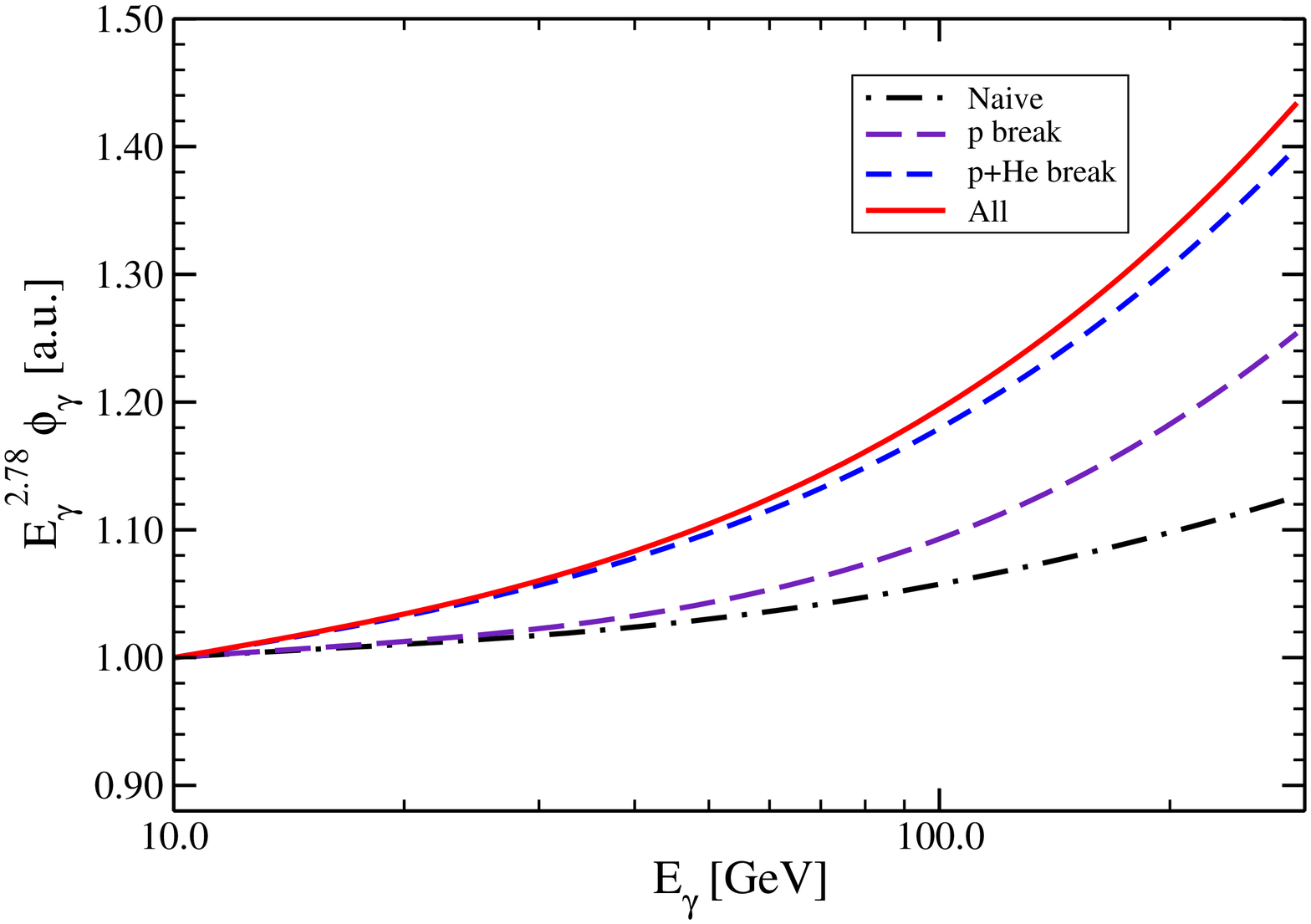} 
\end{tabular}
\end{center}
\caption{$\gamma$-ray spectrum: the standard departure from equality of power-law with parent flux (dot-dashed, black), of adding the hardening in the $p$ spectrum
at TeV scale as suggested by CREAM (long-dashed, purple), of assuming the CREAM hardening for {\it both} $p$ and He (short-dashed, blue), and of including the small effect of other nuclei as well (solid, red).
~\label{fig:2}}
\end{figure}

In Fig.~\ref{fig:2}, we show the result of computing the diffuse gamma ray spectrum (via the kernel provided
in~\cite{Kelner:2006tc}\footnote{Note that we are only interested in the effects that different high-energy CR spectra have on the gamma-ray
spectrum  at $E_\gamma\gg 1\,$GeV, so the simplified formalism presented in~\cite{Kelner:2006tc} and valid in the high-energy regime is
sufficient for our purposes.}) using the AMS-01 spectral fits $\phi_i^L$, 
extrapolated to arbitrarily high energy (long-dashed, black curve). The flux has
been multiplied by $E_\gamma^{2.78}$ to underline the departure from identical power-law behaviour between photons and parent CR due to
production
cross section/multiplicity effects.  Instead, if one keeps $\epsilon_{\rm M}=1.84$, but introduces the broken power-law spectrum for the protons {\it
only} as from Eq.~(\ref{broknP}), around 300 GeV one would obtain $\sim 10\%$ higher gamma fluxes, as shown by the long-dashed, purple curve in
Fig.~\ref{fig:2}. This case is introduced in order to gauge visually the effect of the break of $2.78-2.66\simeq 0.12$ in the
spectral index, between AMS and CREAM determination of proton spectra (to be compared with the $\sim 0.01$ and $0.02$ fit errors, respectively,
reported by the experiments).  The solid, red curve shows the effect of  ``discrepant hardenings'' of the spectra, namely
the $T$ dependence of $\epsilon_{\rm M}$. This constitutes the major distortion and is mostly due to He (as shown by the short-dashed, blue curve); 
overall, the spectrum around 300 GeV is 30\% higher with respect to naive expectations.

The effect discussed here is already ``an estimate of error'': 
assessing the error on this quantity goes beyond our purpose. However, we can 
safely conclude that our results are not significantly affected by the {\it statistical errors} with which the normalizations or spectral indices are known. We have
checked this explicitly as follows: while keeping the Helium fluxes at low and high energies at the values described above, we have varied the
normalization of the $p$ fluxes at low/high energy in such a way  that the $p$ to He ration varies within $\pm 1$ (i.e. twice the
statistical error) of 18.8 at 100 GeV/nucleon \cite{Aguilar:2002ad}, and within $\pm 0.6$ of 8.9 at 9 TeV/nucleon \cite{Ahn:2010gv}. 
The resulting variations in {\it the shape} of
secondary radiation are negligible (i.e. at most at a few percent level). This is due to the fact that the (relatively small) effect of $p$ flux
renormalization  and the change in $\epsilon_{\rm M} $ anti-correlate, and tend to
cancel each other. On the other hand, the exact value of the spectral hardening is more important: 
Fig.~\ref{fig:2} shows that more than 1/3 of the
hardening is due to the assumed ``best fit'' spectral index difference of $\pm 0.12$ between low and high energy. 
This should be compared with
the {\it statistical} errors of about $\sim \pm 0.01$ and $\pm 0.02$ quoted by the AMS and CREAM collaborations, respectively. 

\section{Discussion and Conclusions}~\label{disconcl}
In this article we have argued that departure at high energy from a simple and universal power-law for all cosmic ray
spectra, as suggested by recent data, should cause a spectral distortion in the spectra of secondary cosmic ray
yields (like diffuse photons and antiprotons) compared to the predictions obtained extrapolating the best fits
to low-energy data sets.
We have illustrated this effect using the best fit results of AMS-01 data at low energies and the CREAM data at 
high-energy,
finding effects exceeding $10\%$ above $\sim$100 GeV, and reaching about 30\% for photons around 300 GeV and 
for $\bar{p}$ close
to TeV energy; this figure is somewhat sensitive to the systematic error on  the spectral index at high energy as 
well as other eventual systematics
which do not cancel out in ratios of species (like $p$/He). 
If the hardening in nuclei data would be due to local effect and not representative of the ISM average, 
the effect on the antiproton yield may be small while the hadronic diffuse gamma-ray spectrum could 
be modified differently according to the line-of-sight. Notice that the effect of a possible harder nuclear spectra on atmospheric neutrinos 
was already estimated in~\cite{Barr:2006it}. 

One might wonder how relevant is a high-energy effect of a few tens of percent in a field where data are usually plagued by larger errors. 
We think that, at present, this level of accuracy is  becoming crucial for at least a couple of reasons: 
First, space experiments like FERMI or the future 
AMS-02~\cite{ams02} are introducing us to a new era of large exposures, which can reveal more subtle features than previous
cosmic ray or gamma-ray experiments. Fermi data errors at $E_\gamma\simeq 100\,$GeV are already $\sim \pm 20\%$~\cite{Abdo:2010nz}, 
and forecasts that have been presented suggest that AMS-02 (if performing close to specifications) will be  certainly sensitive to effects of this magnitude, see for  example~\cite{Casaus:2009zz}.
 Second, both diffuse gamma-rays~\cite{Abdo:2009mr} and the combination of hadronic data~\cite{DiBernardo:2009ku}
 are consistent, at least at leading order, with a ``standard'' scenario for the production and propagation of cosmic rays in the Galaxy.
 It is very likely that any departure from baseline models, if detectable, is going to be present at such a sub-leading level. Modelling thus
 the astrophysical background for indirect DM searches as a simple power law, as often done in the literature, might lead to wrong 
 conclusions about the evidence of a signal, or to a bias in the inferred  values of the parameters describing the new phenomena, should
 they be detected. 

Even in a conservative scenario, the detection of such spectral signatures in secondary channels would provide a way to check the {\it
interstellar} nature of the spectral features in the cosmic ray flux at the Earth suggested by the present experiments. We believe that
secondaries provide an important handle for an {\it empirical} cross-check. One should also consider the partial degeneracy of such effects with the
extraction of propagation parameters,  in order to fully exploit the statistical power of forthcoming
data sets. Knowing better the primary flux shapes would allow one to set  strategies minimizing these effects. Last but not least, a multi-messenger
approach would allow one to disentangle these features from alternative sources of spectral distortions: features similar to the ones discussed in this article
arise e.g. in models where high energy $\bar{p}$ are produced in sources~\cite{Blasi:2009bd}, but in that case also associated signatures in secondary/primary
``metals''~\cite{Mertsch:2009ph} (and possibly in high energy neutrinos~\cite{Ahlers:2009ae}) are expected, which are absent for the process described here.

While we are entering a much higher precision era in cosmic ray studies, it is important to keep in mind a couple of points: i) that multi-messenger and multi-channel analyses are mandatory, if one is to gain some deeper knowledge of cosmic ray astrophysics. 
ii) That any hope for the detection of new physics (not to speak of extracting
new physics parameters) requires a more robust understanding of the possible range of astrophysical yields. In that respect, a natural development
of this initial investigation would be to (re)assess how the errors on primary flux knowledge map into the predictions for secondaries (including
their normalization), as much as possible in a parameterization-independent way. \\

\acknowledgments
We warmly acknowledge David Maurin for the agreement to use the
USINE code for the calculation of the antiproton flux.


\begin{thebibliography}{100}

\bibitem{Barr:2006it}
  G.~D.~Barr, T.~K.~Gaisser, S.~Robbins and T.~Stanev,
``Uncertainties in atmospheric neutrino fluxes,''
  Phys.\ Rev.\  D {\bf 74}, 094009 (2006).

\bibitem{Donato:2001ms}
  F.~Donato, D.~Maurin, P.~Salati, A.~Barrau, G.~Boudoul and R.~Taillet,
  ``Antiprotons from spallation of cosmic rays on interstellar matter,''
  Astrophys.\ J.\  {\bf 563}, 172 (2001).

\bibitem{Donato:2008jk}
  F.~Donato, D.~Maurin, P.~Brun, T.~Delahaye and P.~Salati,
  ``Constraints on WIMP Dark Matter from the High Energy PAMELA $\bar{p}/p$ data,''
  Phys.\ Rev.\ Lett.\  {\bf 102}, 071301 (2009).

\bibitem{Blasi:2009bd}
  P.~Blasi and P.~D.~Serpico,
``High-energy antiprotons from old supernova remnants,''
  Phys.\ Rev.\ Lett.\  {\bf 103}, 081103 (2009).

\bibitem{Ahn:2010gv}
  H.~S.~Ahn {\it et al.},
  ``Discrepant hardening observed in cosmic-ray elemental spectra,''
  Astrophys.\ J.\  {\bf 714}, L89 (2010).

\bibitem{Panov:2006kf}
  A.~D.~Panov {\it et al.},
``Elemental energy spectra of cosmic rays from the data of the ATIC-2 experiment,''
Bulletin of the Russian Academy of Sciences: Physics  {\bf 71}, Vol. 04, 494-497  (2007)
[astro-ph/0612377].

\bibitem{adrianitalk}
Talk by O. Adriani at 35th International Conference on High Energy Physics, Paris 2010. Slides available at \texttt{http://pamela.roma2.infn.it}.


\bibitem{Biermann:2010qn}
  P.~L.~Biermann, J.~K.~Becker, J.~Dreyer, A.~Meli, E.~S.~Seo and T.~Stanev,
``The origin of cosmic rays: Explosions of massive stars with magnetic winds
and their supernova mechanism,''
  Astrophys.\ J.\  {\bf 725}, 184 (2010).

\bibitem{Aguilar:2002ad}
  M.~Aguilar {\it et al.}  [AMS Collaboration],
``The Alpha Magnetic Spectrometer (Ams) On The International Space Station.
I: Results From The Test Flight On The Space Shuttle,''
  Phys.\ Rept.\  {\bf 366}, 331 (2002)
  [Erratum-ibid.\  {\bf 380}, 97 (2003)].


  \bibitem{Adriani:2008zr}
  O.~Adriani {\it et al.}  [PAMELA Collaboration],
``An anomalous positron abundance in cosmic rays with energies 1.5-100 GeV,''
  Nature {\bf 458}, 607 (2009).
\bibitem{Abdo:2009zk}
  A.~A.~Abdo {\it et al.}  [Fermi LAT Collaboration],
``Measurement of the Cosmic Ray e+ plus e- spectrum from 20 GeV to 1 TeV with
the Fermi Large Area Telescope,''
  Phys.\ Rev.\ Lett.\  {\bf 102}, 181101 (2009).
  
\bibitem{Ackermann:2010ij}
  M.~Ackermann {\it et al.}  [Fermi LAT Collaboration],
 ``Fermi LAT observations of cosmic-ray electrons from 7 GeV to 1 TeV,''
  Phys.\ Rev.\  D {\bf 82}, 092004 (2010).


\bibitem{Serpico:2008te}
  P.~D.~Serpico,
 ``On the possible causes of a rise with energy of the cosmic ray positron
 fraction,''
  Phys.\ Rev.\  D {\bf 79}, 021302 (2009).

\bibitem{Delahaye:2010ji}
  T.~Delahaye,  J.~Lavalle, R.~Lineros, F.~Donato and N.~Fornengo,
  ``Galactic electrons and positrons at the Earth:new estimate of the primary
  and secondary fluxes,''
  Astron.\ Astrophys.\  {\bf 524}, A51 (2010).


\bibitem{usineI} D. Maurin, F. Donato, R. Taillet, P. Salati, 
``Cosmic rays below z=30 in a diffusion model: new constraints on propagation parameters,''
 Astrophys.\ J.\  {\bf 555}, 585 (2001).


\bibitem{Abdo:2009ka}
  A.~A.~Abdo {\it et al.}  [Fermi LAT Collaboration],
  ``Fermi LAT Observation of Diffuse Gamma-Rays Produced Through Interactions
  between Local Interstellar Matter and High Energy Cosmic Rays,''
  Astrophys.\ J.\  {\bf 703}, 1249 (2009).

\bibitem{Mori:2009te}
  M.~Mori,
  ``Nuclear enhancement factor in calculation of Galactic diffuse gamma-rays: a
  new estimate with DPMJET-3,''
  Astropart.\ Phys.\  {\bf 31}, 341 (2009).
  
  \bibitem{Kelner:2006tc}
  S.~R.~Kelner, F.~A.~Aharonian and V.~V.~Bugayov,
 ``Energy spectra of gamma-rays, electrons and neutrinos produced at  proton
 proton interactions in the very high energy regime,''
  Phys.\ Rev.\  D {\bf 74}, 034018 (2006)
  [Erratum-ibid.\  D {\bf 79}, 039901 (2009)].
  
  \bibitem{ams02}
\texttt{http://www.ams02.org/}

  
\bibitem{Abdo:2010nz}
  A.~A.~Abdo {\it et al.}  [Fermi-LAT collaboration],
 ``The Spectrum of the Isotropic Diffuse Gamma-Ray Emission Derived From
 First-Year Fermi Large Area Telescope Data,''
  Phys.\ Rev.\ Lett.\  {\bf 104}, 101101 (2010).

  \bibitem{Casaus:2009zz}
  J.~Casaus,
 ``The AMS-02 experiment on the ISS,''
  J.\ Phys.\ Conf.\ Ser.\  {\bf 171}, 012045 (2009).

\bibitem{Abdo:2009mr}
  A.~A.~Abdo {\it et al.}  [Fermi LAT Collaboration],
 ``Fermi Large Area Telescope Measurements of the Diffuse Gamma-Ray Emission
 at Intermediate Galactic Latitudes,''
  Phys.\ Rev.\ Lett.\  {\bf 103}, 251101 (2009).

\bibitem{DiBernardo:2009ku}
  G.~Di Bernardo, C.~Evoli, D.~Gaggero, D.~Grasso and L.~Maccione,
 ``Unified interpretation of cosmic-ray nuclei and antiproton recent
 measurements,''
  Astropart.\ Phys.\  {\bf 34}, 274 (2010).

\bibitem{Mertsch:2009ph}
  P.~Mertsch and S.~Sarkar,
  ``Testing astrophysical models for the PAMELA positron excess with cosmic ray
  nuclei,''
  Phys.\ Rev.\ Lett.\  {\bf 103}, 081104 (2009).

\bibitem{Ahlers:2009ae}
  M.~Ahlers, P.~Mertsch and S.~Sarkar,
  ``On cosmic ray acceleration in supernova remnants and the FERMI/PAMELA
  data,''
  Phys.\ Rev.\  D {\bf 80}, 123017 (2009).



\end{thebibliography}
\end{document}